\documentclass[twocolumn,showpacs,amsmath,amssymb]{revtex4}
\usepackage{bm}

\begin{document}
\newcommand{\pp}[1]{\phantom{#1}}
\newcommand{\be}{\begin{eqnarray}}
\newcommand{\ee}{\end{eqnarray}}
\newcommand{\ve}{\varepsilon}
\newcommand{\vs}{\varsigma}
\newcommand{\vp}{\varphi}
\newcommand{\Tr}{{\rm Tr\,}}

\title{
Comment on `Quantum entropy and special relativity' by A. Peres, P. F. Scudo, and D. R. Terno, Phys. Rev. Lett. {\bf 88}, 230402 (2002).
}
\author{Marek Czachor}
\address{
Katedra  Fizyki Teoretycznej i Metod Matematycznych 
\\
 Politechnika Gda\'{n}ska,
ul. Narutowicza 11/12, 80-952 Gda\'{n}sk, Poland
}
\maketitle

Peres, Scudo, and Terno (PST) point out in \cite{PST}  that the entropy of relativistic qubits is not a relativistic scalar. The crucial element of the construction is the reduced density matrix of `spin degrees of freedom' defined by a partial trace over momenta of a full 1-particle density matrix of a massive spin-1/2 particle. I understand that the Authors have in mind an experiment where detectors are blind to the particle momentum and distinguish only different polarizations. In this context I want to point out two elements that have led to some controversies concerning the physical meaning of the phenomenon. 

If the momentum operator has spectral representation $\bm P=\int d\Gamma(\bm p)
\bm p|\bm p\rangle\langle\bm p|$ then the helicity operator $W_0=\bm J\cdot \bm P$ reads
$W_0=\sum_{s=\pm 1/2}s\int d\Gamma(\bm p)\Pi(\bm p,s)|\bm p\rangle\langle\bm p|=\sum_ss\Pi_s$. 
If $u(\bm p,s)$ is an eigenvector of $\Pi(\bm p,s)$ then a general state is 
$|\psi\rangle=\sum_s\int d\Gamma(\bm p)\psi(\bm p,s)u(\bm p,s)|\bm p\rangle$. 
A probability that the particle has helicity $s$ is given by 
$p_s=\langle \psi|\Pi_s|\psi\rangle=\int d\Gamma(\bm p)|\psi(\bm p,s)|^2$. Elliptic polarizations correspond to observables $UW_0U^{\dag}$ where $U$ may be momentum dependent. 
If $U$ is momentum independent then the probabilities associated with $U\Pi_sU^{\dag}$ can be computed in terms of the PST density matrix 
$\rho_{ss'}=\int d\Gamma(\bm p)\psi(\bm p,s)\overline{\psi(\bm p,s')}$ and certain new `reduced' projectors $\pi_r$. 

The bad news is that observables with momentum-dependent $U$ need new PST-type density matrices. For this reason the PST reduced state does not describe all the possible polarization experiments, even if measurements are blind to momenta. An interesting class of such momentum-dependent polarizations is produced by Lorentz boosts. 
Unitary representations of the Poincar\'e group map $W_0$ into 
$\sum_{s=\pm 1/2}s\int d\Gamma(\bm p)D(\bm p)\Pi(\bm p,s)
D(\bm p)^{\dag}|\bm{\Lambda^{-1}p}\rangle\langle\bm{\Lambda^{-1}p}|$. $D(\bm p)$ is unitary. For $m=0$ it commutes with $\Pi(\bm p,s)$ and depends on $\bm p$ only via $\bm p/|\bm p|$. Lorentz transformations change momenta (Doppler effect) and map circular polarizations into combinations of linear polarizations (unless $m=0$) with different polarization planes for different momenta. This is why the PST $\rho_{ss'}$ changes its entropy: Some information escapes into polarizations that have been tossed out with the bathwater of the traced-out 
`momentum bath'. There is no problem if one works with the full state 
$\rho_{ss'}(\bm p,\bm p')$. This full state may be regarded as collection of all the possible reduced PST states. The PST analysis provides us with an interesting argument supporting earlier claims that relativistic qubits are inherently momentum dependent, and thus one should expect a kind of relativistic noise \cite{MC-PRA,SPIE}. 

Now, the good news is that if instead of $W_0$ one takes $t_\mu W^\mu$ with
$t_\mu$ chosen to be an eigenvector of $\Lambda$, then  the unitary matrix 
$D(\bm p)$ is momentum independent \cite{MC,MCMW}. Accordingly,
the entropy of the PST density matrix is unchanged by the boost. For the same reason there is no change of entanglement, in the PST sense, between appropriate qubits in multiparticle states if the qubits correspond to projections of $W_\mu$ in an eigendirection of the Lorentz transformation. Eigenvectors of boosts are {\it null\/} (i.e. $t_\mu t^\mu=0$) and it is an experimental challenge how to measure such observables. A hint that may help is that the 
`gauge transformation' $t'_\mu= t_\mu+\theta P_\mu$ does not change the product 
$t_\mu W^\mu$ since $P_\mu W^\mu=0$, and thus the resulting eigenvectors are unchanged. For example, the null $t_\mu$ can be replaced by a spacelike, momentum dependent direction with
$\theta=-t_\mu P^\mu/m^2$, 
$t'_\mu P^\mu=0$. It can be guaranteed that a given $\rho_{rs}$ can be written in terms of such PST-type qubits that its entropy will be left unchanged by a given boost. In principle, there exist an error-correcting procedure that allows to eliminate the leak-out of information from the PST state: For any motion one has to appropriately adjust the spin quantization axis. One may regard it as a sort of relativistic analogue of a decoherence-free subspace.

My work on relativistic aspects of qubits was supported by the KBN Grant No. PBZ-Min-008/P03/2003.

\end{document}